\documentclass[conference]{IEEEtran}

\IEEEoverridecommandlockouts

\usepackage{cite}
\usepackage{amsmath,amssymb,amsfonts}
\usepackage{graphicx}
\usepackage{textcomp}
\usepackage{xcolor}
\def\BibTeX{{\rm B\kern-.05em{\sc i\kern-.025em b}\kern-.08em
    T\kern-.1667em\lower.7ex\hbox{E}\kern-.125emX}}

\usepackage{algorithm}
\usepackage{algpseudocode}

\usepackage{listings}
\makeatletter
\def\lst@makecaption{%
  \def\@captype{table}%
  \@makecaption
}
\makeatother

\AtBeginDocument{%

}

\usepackage{comment}

\usepackage[acronym]{glossaries}

\newacronym{gpt}{GPT}{Generative Pre-trained Transformer}
\newacronym{llm}{LLM}{large language model}
\newacronym{rope}{RoPE}{Rotary Position Embedding}
\newacronym{sc}{SC}{smart contract}
\newacronym{clm}{CLM}{Casual Language Modeling}
\newacronym{acc}{ACC}{accuracy}
\newacronym{rl}{RL}{reinforcement learning}
\newacronym{rlhf}{RLHF}{reinforcement learning from human feedback}

\usepackage{booktabs}
\usepackage{tabularx,ragged2e}
\usepackage{threeparttable}
\usepackage{placeins}
\usepackage{cleveref}
\usepackage{pgfplots, pgfplotstable}

\usepackage{layouts}
\usepackage{url}

\def\citet{\cite}

\makeatletter
\newcommand{\linebreakand}{%
  \end{@IEEEauthorhalign}
  \hfill\mbox{}\par
  \mbox{}\hfill\begin{@IEEEauthorhalign}
}
\makeatother

\usepackage{fancyhdr}

\fancypagestyle{copyrightnotice}{
    \fancyhf{}%

    \fancyhead[L]{\parbox[b]{\textwidth}{\footnotesize Accepted to the 34th IEEE International Symposium on Software Reliability Engineering (ISSRE 2023), October 9--12, 2023, Florence, Italy}}%
    \fancyfoot[L]{\parbox[t]{\textwidth}{\footnotesize \textcopyright{} 2023 IEEE. Personal use of this material is permitted. Permission from IEEE must be obtained for all other uses, in any current or future media, including reprinting/republishing this material for advertising or promotional purposes, creating new collective works, for resale or redistribution to servers or lists, or reuse of any copyrighted component of this work in other works.}}%
}

\begin{document}

\title{Efficient Avoidance of Vulnerabilities in Auto-completed Smart Contract Code Using Vulnerability-constrained Decoding}

\author{\IEEEauthorblockN{Andr\'e Storhaug\textsuperscript{\dag}, Jingyue Li\textsuperscript{\dag}, Tianyuan Hu\textsuperscript{\ddag}}

\dag \textit{Department of Computer Science, Norwegian University of Science and Technology, Trondheim, Norway}\\
\ddag \textit{School of Computer Science and Engineering, Southeast University, Nanjing, China} \\
andre.storhaug@ntnu.no, jingyue.li@ntnu.no, tianyuanhu@seu.edu.cn}
\maketitle

\thispagestyle{copyrightnotice}

\begin{abstract}
Auto-completing code enables developers to speed up coding significantly. Recent advances in transformer-based large language model (LLM) technologies have been applied to code synthesis. However, studies show that many of such synthesized codes contain vulnerabilities. We propose a novel vulnerability-constrained decoding approach to reduce the amount of vulnerable code generated by such models. Using a small dataset of labeled vulnerable lines of code, we fine-tune an LLM to include vulnerability labels when generating code, acting as an embedded classifier. Then, during decoding, we deny the model to generate these labels to avoid generating vulnerable code. To evaluate the method, we chose to automatically complete Ethereum Blockchain smart contracts (SCs) as the case study due to the strict requirements of SC security. We first fine-tuned the 6-billion-parameter GPT-J model using 186,397 Ethereum SCs after removing the duplication from 2,217,692 SCs. The fine-tuning took more than one week using ten GPUs. The results showed that our fine-tuned model could synthesize SCs with an average BLEU (BiLingual Evaluation Understudy) score of 0.557. However, many codes in the auto-completed SCs were vulnerable. Using the code before the vulnerable line of 176 SCs containing different types of vulnerabilities to auto-complete the code, we found that more than 70\% of the auto-completed codes were insecure. Thus, we further fine-tuned the model on other 941 vulnerable SCs containing the same types of vulnerabilities and applied vulnerability-constrained decoding. The fine-tuning took only one hour with four GPUs. We then auto-completed the 176 SCs again and found that our approach could identify 62\% of the code to be generated as vulnerable and avoid generating 67\% of them, indicating the approach could efficiently and effectively avoid vulnerabilities in the auto-completed code. 
\end{abstract}


\section{Introduction}
Programs can be synthesized by leveraging the semantics of code \cite{alon2018code2vec} or neural networks \cite{iyer2018mapping}. Inspired by the success of large natural language models, e.g., Embeddings from Language Models (ELMo) \cite{peters2018deep} and \acrfull{gpt} \cite{radford2018improving}, large-scale Transformer models have been applied in the domains of code synthesis \cite{copilot, alphacode, christopoulou2022pangu}. 

One kind of code synthesis is to generate complete function-level code based on specifications \cite{copilot,alphacode}. Another type of code synthesis uses existing code as context and developers' comments as input to generate new code to auto-complete a function, e.g.,   using the neural network model bidirectional LSTM (BiLSTM) to automatically propose new code to developers based on existing code and comments specifying the code to be generated \citet{iyer2018mapping}. GitHub Copilot \cite{copilot}, which is based on Codex \cite{chen2021codex} by OpenAI, auto-completes the code to speed up the coding significantly. Massive LLM like ChatGPT \cite{chatgpt} has also enabled the capability to perform program synthesis.

Although state-of-the-art models provide powerful synthesis capabilities, they also have challenges \cite{derner2023safeguards}, especially related to security. Pearce et al. \citet{pearce2021asleep}  analyzed the insecure code generated by GitHub Copilot. Their analysis of 1,689 synthesized Python and C programs concluded that approximately 40\% of the synthesized code was vulnerable, showing a dire need to reduce the number of vulnerabilities generated with language models. The much larger ChatGPT model also faces many of the same issues. Khoury et al. \cite{khoury2023secure} evaluated 21 generated programs in different programming languages and found that 16 were vulnerable.

To eliminate vulnerabilities in the generated code, once the vulnerabilities in the training datasets are identified, the ideal approach is to remove the vulnerable code in the training dataset and re-train the whole model. However, re-training the whole model can be slow. For a popular LLM used by many developers daily to generate code, its slow re-training may result in many vulnerabilities being introduced in the auto-completed codes. Thus, it is critical to be able to quickly fine-tune the model to reduce the impact of the vulnerabilities. Facing such a challenge, this study aims to answer the research question: \textbf{How to efficiently fine-tune the transformer model to avoid vulnerabilities in the auto-complete code}? To answer the research question, we propose the vulnerability-constrained decoding method, which fine-tunes the transformer using labeled vulnerable code and then restricts the model from generating vulnerable code by using labels as guidance during the decoding. 

To evaluate our method, we performed case studies focused on synthesizing SCs, which have an exceptionally high demand for security, as SC codes are very difficult to be changed after the contracts are deployed. SCs are also desirable targets for adversaries due to the monetary and anonymous nature of Blockchain \cite{atzei2017survey}. We first downloaded Ethereum SCs and fine-tuned the pre-trained GPT-J-6B model \cite{gpt-j} to auto-complete SCs with a BLEU score of 0.557. ``\textit{A BLEU score of 0.6 or 0.7 is considered the best you can achieve \cite{Bleuscore}.}'' The results show that our model can auto-complete SCs with the desired functions. We hypothesized that many of the generated functions might be vulnerable because we used real Ethereum SCs, which might contain vulnerabilities, as the model training dataset. We chose 176 SCs containing ten types of vulnerabilities from the \cite{Hu2023} dataset and found out that over 70\% of the auto-completed SCs based on the code before the vulnerable lines are insecure. After implementing the vulnerability-constrained decoding approach, we re-completed the same 176 SCs. Compared to re-training the who model, which needs more than one week with ten GPUs, the fine-tuning to implement the vulnerability-constrained decoding approach takes only one hour with four GPUs and could avoid many of the vulnerabilities. Our contributions are as follows:
\begin{itemize}
    \item We propose a novel vulnerability-constrained decoding approach to avoid vulnerabilities in the generated code efficiently and effectively. The approach can help companies quickly update \acrshortpl{llm} to avoid the negative consequence caused by vulnerabilities in the model training dataset. 
    \item To the best of our knowledge, our study is the first one using a fine-tuned transformer-based language model to auto-complete SCs with a high BLEU score.
    \item Last but not least, we have created the current largest SC dataset that could be used to train transformers to auto-complete SCs.

\end{itemize}

The rest of the paper is organized as follows. We explain the design of our approach in \Cref{chap:methodology}. \Cref{chap:implementation-evaluation} presents the implementation of the approach and the evaluations. \Cref{chap:related-work} introduces related work. \Cref{chap:discussion} discusses our results. \Cref{chap:conclusion} concludes the study and proposes future work.

\section{Our Approach}
\label{chap:methodology}

In general, there are two strategies to auto-complete secure code. One is to generate secure code in the first place from the model, and another is to post-process the generated code using code analysis tools to remove the generated vulnerabilities. One critical challenge of the post-processing strategy is that vulnerability filtering can be slow because the type of vulnerability in the generated code is usually unknown. Therefore, scanning the generated code for all possible types of vulnerabilities is necessary. As developers usually expect real-time code auto-completion, heavy scans will slow down the code generation performance and negatively impact the developers' user experience. Another challenge is that most vulnerability detectors, such as Oyente \cite{oyente2016making}, Slither \cite{feist2019slither}, and Mythril \cite{mythril}, require complete functions as input to scan SCs. Many dynamic detectors, such as ContractFuzzer \cite{ContractFuzzer}, sFuzz \cite{sFuzz}, and SMARTIAN \cite{SMARTIAN} to analyze SCs, need bytecode as input that must be compiled from the complete source code. When auto-completing code using an LLM, the codes are, most of the time, incomplete functions. Thus, we choose to focus on the first strategy. 

As aforementioned, the ideal approach is to train the code-generation model with only secure code. However, this approach may not apply in practice because people may find vulnerabilities in the training dataset after the model has been trained and in operation. To reflect the real-world scenario, we simulate the scenario that a model is trained based on available code and that the vulnerabilities in the training dataset are found after the model is trained. Our general idea is to let the model recognize vulnerable code, which is then used to steer the model to avoid generating insecure code. This study contains three steps to investigate the idea.

\textbf{Step 1: Collect the source code data and train an LLM to auto-complete code.}
This step aims to generate a transformer model to auto-complete code using the available code as training, validation, and test datasets. 

\textbf{Step 2: Collect and label vulnerable code and update the model.} 
Although we expect that the model trained from step 1 can auto-complete code to provide the desired functionalities, the generated code may contain vulnerabilities because not all codes in the training dataset are secure. We assume that people have identified and reported vulnerabilities after the model is trained and that the vulnerability information, including vulnerability type and location of the vulnerable lines, is available. We use the vulnerability information to label the vulnerable codes and apply the labeled codes to fast fine-tune the model trained from step 1. 

\textbf{Step 3: Restrict the model from generating vulnerable code.} 
When auto-completing code, we believe it is better not to generate code than to generate vulnerable code. This step aims to force the fine-tuned model to generate only secure code by denying the model to generate the vulnerability labels inserted into the model through fine-tuning in step 2 to avoid  generating vulnerable code. As mentioned, we use SCs and their vulnerabilities as case studies to explain and evaluate our approach.

\subsection{Step 1: Collect source code data and train an LLM to auto-complete code}
\subsubsection{Smart contract collection}
\label{sec:smart-contract-collection}
We follow mostly the same SC collection method as \cite{Hu2023}. We use Google BigQuery to select all SC addresses available on the Ethereum Blockchain with at least one transaction. We then query Etherscan, the largest provider of verified SCs, for the SCs (source code) at these addresses. Following \cite{Hu2023}, we also filter the contracts for uniqueness using the computationally efficient token-based similarity algorithm Jacard Index \cite{allamanis_adverse_2019}. The downloaded SCs contain many duplicated library codes. To mitigate this, we ``inflate'' the contracts, meaning each contract file is split into its original representative files. The library code is split (read inflated) into separate contract records along with other imported contract files. The resulting SCs were then grouped by file name and filtered for uniqueness with a similarity threshold of 0.9, calculated using the Jacard Index \cite{jaccard}, which means that all SCs whose code shares more than 90\% of the tokens are discarded.

\subsubsection{Training transformer model for SC auto-completion}
\label{sec:language-model}
We use an autoregressive transformer language model to auto-complete the SC code. The goal is to fit a model \(f: \mathbb{R}^{N \times D} \rightarrow \mathbb{R}^{N \times D}\), where \(N\) is the sequence length of tokens of dimension \(D\). An autoregressive model tries to estimate the conditional probability distribution, \(p\):

\begin{equation}
p(x_{1:T} \mid c) = \prod_{t=1}^{T} p(x_t \mid x_{1:t-1},c)
\end{equation}

\noindent where \(x_t \in \mathbb{R}^{D}\) is the \(t\)’th observation of a discrete token, \(c\) is the initial context (input) sequence, and \(T\) is the length of the sequence of predicted tokens.

\begin{figure}[htbp]
    \centering
    \includegraphics[width=\columnwidth]{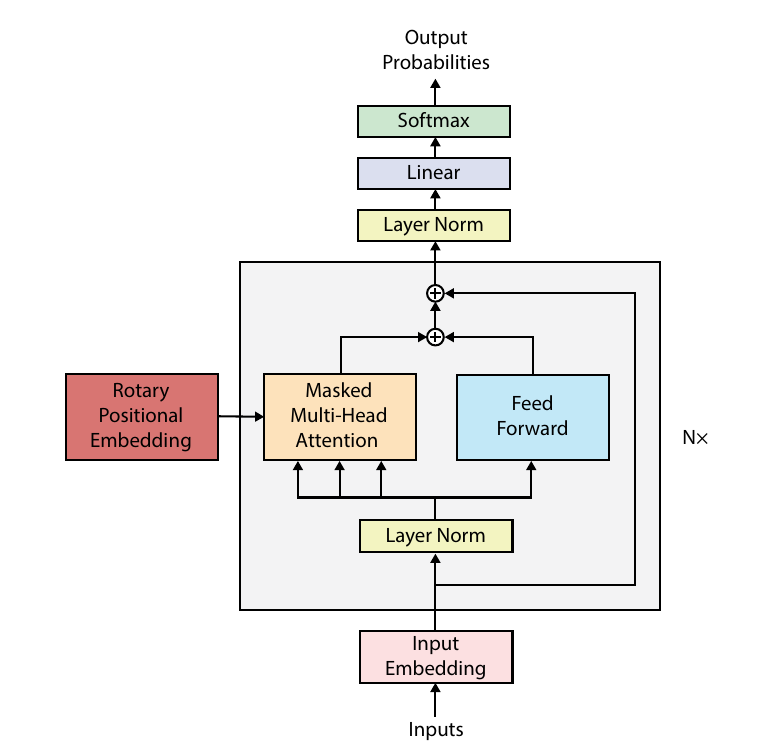}
    \caption{Diagram of GPT-J model architecture.}
    \label{fig:gpt-j-architecture}
\end{figure}

Several auto-regressive transformer models, e.g., \cite{chen2021codex} and ChatGPT \cite{chatgpt}, are described in the literature. However, only a few of them have open-sourced pre-trained weights. Among the open-sourced models, GPT-J \cite{gpt-j} is one of the popular large models that includes code in its pre-training dataset ``The Pile'' \cite{gao2021thepile}, which is an 825 GB open-source dataset. GPT-J was released in June 2021 by EleutherAI \cite{eleutherai} and outperformed existing open-source \acrshort{gpt} systems in qualitative programming evaluations \cite{wolf2021}. GPT-J uses a decoder-only architecture, as shown in \Cref{fig:gpt-j-architecture}. GPT-J introduces some notable differences from standard transformer models. Firstly, instead of computing attention and feed-forward layers in sequential order, they are computed in parallel, and the results are added together, which decreases communication during distributed training, resulting in increased throughput. Secondly, GPT-J uses \acrfull{rope} \cite{su2021roformer} for position encoding, which is shown to result in better model quality in tasks with long text \cite{su2021roformer}. So, we chose to use GPT-J-6B, which is a GPT-J model with six billion parameters, as the pre-trained model and fine-tune it using the SCs we collected. 

\subsection{Step 2: Collect and label vulnerable code for vulnerability-tuning the model}
\label{sec:step2}
\subsubsection{Collect and label vulnerabilities data}
\label{sec:vulnerable-contracts}

Our labeled dataset is based on the dataset \footnote{\url{https://doi.org/10.5281/zenodo.7744053}} in \cite{Hu2023}, which contains manually labeled real-world vulnerable \acrshortpl{sc}. The dataset covers ten types of SC vulnerabilities. The dataset was acquired by running several SC code analysis tools to identify the vulnerabilities in 136,969 \acrshortpl{sc}. The authors then manually verified the resulting vulnerabilities to ensure they were vulnerable and exploitable. Although several other datasets, e.g., the dataset provided by \cite{solidifi} and  \cite{EmpiricalOnSmartBug}, are available, we believe the SCs in the dataset of \cite{Hu2023} are more reliable and contain more critical vulnerabilities because the focuses of \cite{Hu2023} is to eliminate false alarms of SC vulnerability detectors and focus on vulnerabilities that urgent to be fixed, i.e., the exploitable ones.

The dataset of \cite{Hu2023} does not provide the location of the vulnerabilities within a contract. We, therefore, reached out to the authors and got a copy of the SCs labeled by the different tools. Following \cite{Hu2023}, we only care about the contracts flagged as vulnerable by multiple tools. Since we need the actual line location of the vulnerabilities, we calculate the overlapping lines of the vulnerabilities within each SC. Different tools report slightly different results of the actual location of the vulnerabilities. To account for this, we allow some distance between the reported lines. We set this distance to \(\pm\) five lines. If multiple vulnerabilities occur within these ten lines, there will be duplications. We, therefore, filter the results so that each tool's reported vulnerability can only be used in one overlap combination while also selecting the combination that gives the minimum distance between the reported lines.

After calculating the overlapping vulnerabilities, we transfer these onto our inflated dataset, as described in \cref{sec:smart-contract-collection}. We then extract these contracts based on the contract addresses of the SCs in the dataset of \cite{Hu2023}. We get 609 vulnerable contracts containing 1,117  vulnerabilities. One vulnerable contract may contain several vulnerabilities.

\subsubsection{Using labeled vulnerable code for vulnerability-tuning}
\label{sec:labeling}

Starting from the fine-tuned model from step 1, we use the labeled vulnerable SC dataset for the model's future fine-tuning (called vulnerability-tuning from here). The goal is to train the model to generate vulnerability labels at the start of lines containing vulnerable code. Hence, the model learns to associate a particular label with a certain type of vulnerability. We prepend the vulnerable lines for each vulnerability in the dataset with the vulnerability label \(V_i\) corresponding to an abbreviation of its vulnerability type, e.g., IOU (meaning Integer Overflow or Underflow). Note that the labels will be inserted after whitespace indentation and before code statements. \Cref{lst:example-labeling} shows an example result after injecting an IOU label, marked in green.

\begin{lstlisting}[
    caption={Example of labeling a vulnerable line.},
    label=lst:example-labeling,
    language=Solidity,
    belowskip=0pt,
    escapechar=§]
function increaseLockTime(uint _seconds) public {
    §\lstbg{green!40}{<IOU>}§lockTime[msg.sender] += _seconds;
}
\end{lstlisting}

\subsection{Step 3: Vulnerability-constrained decoding}
To restrict the model during decoding (generation), we provide the model with a list of forbidden tokens (labels). Given a set of \(N\) vulnerability labels (tokens) \(V_i\), i.e., the abbreviations of different types of vulnerabilities, we manipulate the probability \(p\) for selecting any one of these during decoding, such that: \(p(x=V_i)=0\). 

\begin{figure}[htp]
    \centering
    \includegraphics[width=\columnwidth]{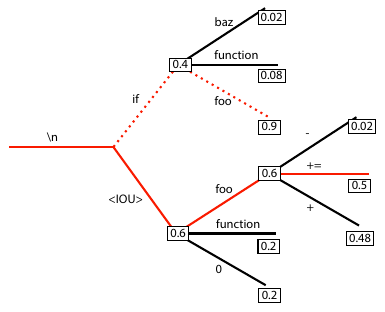}
    \caption{Vulnerability-constrained decoding example using greedy search.}
    \label{fig:greedy_search}
\end{figure}

\Cref{fig:greedy_search} shows an example of the vulnerability-constrained decoding method when using greedy search as the decoding strategy. Greedy search selects the token (word) with the highest probability as its next token: \(x_t = argmax_xP(x \mid x_{1:t-1}) \). Starting from the newline character to the left in the figure, a greedy search would normally have followed the solid red line. Assuming this path would lead to an IOU vulnerability, the model should have labeled this potential vulnerability with an \textless IOU\textgreater token (see \cref{sec:labeling}). If we apply the vulnerability-constrained decoding method, the decoding will avoid this path. It will select the second most probable path instead, following the stippled red line and avoiding the vulnerability altogether. Using a greedy search for decoding strategy is not optimal, as it will miss high-probability words disguised behind low-probability words. However, it is easy to understand and interpret the results. Our method is not limited to greedy decoding and is compatible with most other popular decoding strategies, such as Beam Search \cite{Freitag_2017}.

\section{Implementation and Evaluation}
\label{chap:implementation-evaluation}

This section presents the detailed implementation of each step of our approach and the evaluation results.

\subsection{Smart contract code generation}

\begin{table}[htp]
\centering
\caption{Summary of results.}\label{tab:eval-summary}
\begin{threeparttable}
    \def\arraystretch{1.5}
    \setlength\tabcolsep{2pt} %
    
    \begin{tabular}{cccp{2pt}cc}%
        \toprule
        \textbf{Model} & \multicolumn{2}{c}{\textbf{BLEU score}} & & \multicolumn{2}{c}{\textbf{CrystalBLEU score}}\\\cline{2-3}\cline{5-6}
         & \textbf{Whitespace} & \textbf{Solidity lexer} & & \textbf{Whitespace} & \textbf{Solidity lexer} \\
        \midrule
        \textbf{Pre-trained} & 0.258 & 0.414 & & 0.188 & 0.291\\
        \textbf{Fine-tuned} & 0.557 & 0.557 & & 0.481 & 0.481 \\

       \bottomrule
    \end{tabular}
\end{threeparttable}
\end{table}

\subsubsection{Collected Unique Smart Contracts}
We retrieved 5,810,042 SC addresses from Google BigQuery by the 1st of April 2022. From these addresses, we successfully retrieved 2,217,692 verified SCs from Etherscan. The 2,217,692 SCs are inflated to 5,403,136 separate contract files. After filtering for uniqueness, 5,216,739 duplications are found, giving a duplication percentage of 96.56\%. The resulting Verified Smart Contracts dataset consists of 186,397 SCs. 

\subsubsection{Smart contract code generation model}
\label{sec:smart-contract-generation-model}

We used our collected SCs as the training dataset to fine-tune the pre-trained GPT-J-6B model for generating SC code. To ensure the validity of the model's performance, 80\% of the data were used for training, 10\% for validation, and 10\% for testing. No hyperparameter optimization was performed.

Before training, we randomly shuffled the dataset and fed a complete SC all at once to train the model. For running the training process, the \acrshort{clm} script \footnote{\url{https://github.com/huggingface/transformers/blob/v4.19.0/examples/pytorch/language-modeling/run_clm.py}} provided by HuggingFace was used. Due to the large size of the GPT-J-6B model, we used the deep learning optimization library DeepSpeed \cite{deepspeed} as a wrapper around the HuggingFace library. The training process runs for two epochs. At every five steps, the model is evaluated on 256 samples from the validation split of the SC dataset. Using \textbf{ten NVIDIA A100 40GB GPUs}, the training is completed after \textbf{seven days and four hours}. After completion of the training, the model is evaluated on the entire validation split, achieving an \acrfull{acc} of 0.917 and perplexity of 1.510.

\subsubsection{Evaluate the functionality of auto-completed SCs}
\label{sec:eval-quality-synthesis}
When evaluating code synthesizers, we hope the synthesized codes are similar to code that developers may input. Thus, we first measure \textbf{how well the model provides the desired functionality.} 

To answer this question, we used the original code developed by SC developers as the ground truth. We measured the similarity between the synthesized lines of code and the ground truth. The evaluation consists of three steps:
\begin{enumerate}
    \item We randomly selected SCs from the testing data.
    \item We split each of the selected SCs into two parts. The first part of each selected SC, which included code as the contexts and the comments, was used as the input to the model, while the second part (original code) was used as the ground truth.
    \item The generated lines of code were then compared to the ground truth to calculate their similarities.
\end{enumerate}

\Cref{lst:contract-parts} demonstrates how the different parts of a contract are used in the evaluation. Lines 1-11 are the code context, while lines 13-14 are the comments for the code synthesizer to generate code. The target codes (i.e., ground truth) are in lines 15-20. All consecutive lines of code are discarded. In \Cref{lst:contract-parts}, 
this would only be line 21. Since the model is auto-regressive, a custom stopping strategy based on matching braces is implemented for generating well-formed functions.

\begin{lstlisting}[
    caption={Different contract parts.},
    label=lst:contract-parts,
    language=Solidity,
    belowskip=0pt,
    escapechar=`]
// SPDX-License-Identifier: GPL-3.0
pragma solidity >= 0.7.0;

contract Coin {
    // Sends an amount of newly created coins to an address
    // Can only be called by the contract creator
    function mint(address receiver, uint amount) public {
        require(msg.sender == minter);
        require(amount < 1e60);
        balances[receiver] += amount;
    }
    
\end{lstlisting}
\begin{lstlisting}[
    frame=bottomline,
    firstnumber=13,
    aboveskip=0pt,
    language=Solidity,
    escapechar=`
    ]
    // Sends an amount of existing coins
    // from any caller to an address
\end{lstlisting}
\begin{lstlisting}[
    frame=bottomline,
    firstnumber=15,
    aboveskip=0pt,
    language=Solidity,
    escapechar=`
    ]
    function send(address receiver, uint amount) public {
        require(amount <= balances[msg.sender], "Insufficient balance.");
        balances[msg.sender] -= amount;
        balances[receiver] += amount;
        emit Sent(msg.sender, receiver, amount);
    }
\end{lstlisting}
\begin{lstlisting}[
    frame=bottomline,
    firstnumber=21,
    aboveskip=0pt,
    language=Solidity,
    escapechar=`
    ]
}
\end{lstlisting}

The BLEU score compares the generated code to ground truth. BLEU is a commonly used evaluation metric in code synthesis \cite{ren2020codebleu}. However, as BLEU was originally designed for evaluating natural language, it has shortcomings when applied to evaluate code synthesizers because it neglects important syntactic and semantic aspects of codes \cite{ren2020codebleu}. Due to this limitation, adaptations, e.g., CodeBLEU by \cite{ren2020codebleu}, have emerged, incorporating Abstract Syntex Tree (AST) and data-flow analysis. However, we could not find a readily available implementation of CodeBLEU, especially for SC. CrystalBLEU \cite{CrystalBLEU} is a recently (in 2022) proposed metric to compare code similarity. Thus, we decide to measure the code similarity using both BLEU and CrystalBLEU. To use BLEU or CrystalBLEU to measure code similarity, we must configure the tokenization approach. We can for example tokenize using whitespace or a lexer for the programming language, e.g., Solidity.

We randomly selected 10,000 SC samples from the test split of the Verified Smart Contract Code Comments dataset. Each drawn sample contains function ``code, comment'' pairs, and the complete contract code. The contract code is cut at the end of the sampled function comment and fed into the model as input to generate code. The BLEU and CrystalBLEU scores are calculated by comparing the generated code against the ground truth.  We performed the evaluation procedure for the pre-trained and fine-tuned models. The BLEU results using whitespace as the splitter are shown in Figures \ref{fig:performance-code-context_pretrained} and \ref{fig:performance-code-context_finetuned}. On average, the pre-trained model's BLEU is only 0.258, and the fine-tuned transformer achieves a BLEU score of 0.557. This is over a 100\% improvement from the pre-trained model. To calculate CrystalBLEU, we need to configure some parameters: K (trivially shared n-grams samples) and maxN (n-grams from 1 to maxN). This study configures their values as 500 and four, respectively. The CrystalBLEU results using whitespace as the splitter are 0.188 and 0.481 for the pre-trained and the fine-tuned models, respectively. The results also show over a 100\% improvement from the pre-trained model. By using a lexer for the Solidity programming language to tokenize, the BLEU score increased from 0.414 to 0.557 after transformer fine-tuning, and the CrystalBLEU increased from 0.291 to 0.481. The results are summarized in Table \ref{tab:eval-summary}.

\begin{figure}[htp]
    \centering
    \input{figures/context_pretrained_performance_histogram.pgf}
    \caption{BLEU score frequency distribution of 10,000 generated functions with pre-trained model.}
    \label{fig:performance-code-context_pretrained}
    \input{figures/context_finetuned_performance_histogram.pgf}
    \caption{BLEU score frequency distribution of 10,000 generated functions with the fine-tuned model.}
    \label{fig:performance-code-context_finetuned}
\end{figure}

\subsubsection{Evaluate the security of auto-completed SCs}
\label{sec:eval-security-synthesis}
We hypothesized that many of the auto-completed codes could be insecure, similar to what had been discovered in \cite{chen2021codex}. To test this hypothesis, we randomly chose 20\% of each vulnerability type from 1,117 vulnerabilities extracted from the \cite{Hu2023} dataset. The numbers of each type of vulnerability are shown in the second row of \Cref{tab:vulnerability-distribution}. For the rest 80\% of the data of each type of vulnerability, which are also shown in the first row of Table \ref{tab:vulnerability-distribution}, we use them for vulnerability-tuning the model to avoid generating vulnerable code, which is explained in sections \ref{sec:vulnerablity-labeling} and \ref{sec:vulnerability-avoidance}. 

\begin{table}[htp]
\centering
\begin{threeparttable}
    \def\arraystretch{1.5}
    \setlength\tabcolsep{2pt} %
    \caption{Distribution of the vulnerable data for testing the results and avoiding vulnerability.}
    \label{tab:vulnerability-distribution}
    \begin{tabular}{*{11}{c}}%
        \toprule
        \textbf{Split} & \textbf{DC} & \textbf{FE} & \textbf{IOU} & \textbf{NC} & \textbf{RE} & \textbf{TD} & \textbf{TO} & \textbf{TOD} & \textbf{UcC} & \textbf{UpS}\\
        \midrule
        \textbf{Train} & 94 & 2 & 244 & 111 & 56 & 168 & 10 & 131 & 46 & 79 \\
        \textbf{Test} & 36 & 0 & 45 & 20 & 3 & 24 & 5 & 28 & 4 & 11 \\

       \bottomrule
    \end{tabular}
\end{threeparttable}
\end{table}

The brief explanations of the ten types of vulnerabilities are as follows. More detailed explanations are in \cite{Hu2023}.

\textbf{DC (DelegateCall).}
The \textit{address.delegatecall()} function allows a SC to dynamically load external contracts from \textit{address} at runtime. If the attacker can control the external contract and affect the current contract status, the contract is vulnerable to DC.

\textbf{IOU (Arithmetic/Integer Overflow and Underflow).}
An arithmetic overflow or underflow, often called Integer Overflow or Underflow (IOU), occurs when an arithmetic operation attempts to create a numeric variable value that is larger than the maximum value or smaller than the minimum value of the variable
type. If the arithmetic operation may pass a variable type’s maximum or minimum value and  is performed without using SafeMath \cite{safemath}, the contract is vulnerable to IOU.

\textbf{NC (Nested Call).}
The function containing the loop has a high risk of exceeding its gas limitation and triggering an out-of-gas error. If the attacker can control the loop iteration and causes the out-of-gas error, the contract is vulnerable to NC, 

\textbf{RE (Reentrancy).}
The contract vulnerable to RE uses the
\textit{call()} function to transfer ether to an external contract. The external contract can reenter the vulnerable contract by fallback function. If the state variable changes after the \textit{call()} function, the reentrance will cause status inconsistency. 

\textbf{TD (Timestamp Dependency).}
The contract uses the \textit{timestamp} as the deciding factor for critical operations, e.g., sending ether. If the attacker can get ether from the contract by manipulating the \textit{timestamp} or affecting the critical operations, the contract is vulnerable to TD.

\textbf{TO (TxOrigin).}
If the contract only uses \textit{tx.origin} to verify the caller's identification for critical operations, it is vulnerable to TO.

\textbf{TOD (Transaction Order Dependency).}
The contract may send out ether differently according to different values of a global state variable or different balance values of the contract. If the attackers can get ether from the contract by manipulating the transaction sequences, the contract is vulnerable to TOD.

\textbf{UcC (Unchecked Call).}
The contract uses the function \textit{call()} or \textit{send()} without result checking. If the \textit{send()} or \textit{call()}  function fails and leads to status inconsistency, the contract is vulnerable to UcC.

\textbf{UpS (Unprotected Suicide).}
If an attacker can self-destruct the contract by calling the \textit{selfdestruct}(address) function, the contract is vulnerable to UpS. 

\textbf{FE (Frozen Ether).}
If the contract can receive ether but cannot transfer it by itself, it is vulnerable to FE. 

We split the contract above the vulnerable line for each of the 176 vulnerable ones shown in the second row of \Cref{tab:vulnerability-distribution}. We then auto-complete the function using the model and manually investigate whether the generated code contains the original vulnerability type. We do not consider vulnerabilities other than the current sample's original vulnerability type. Further, we only consider the immediate scope of the originally labeled vulnerability line. Among the 176 testing data samples, 50 auto-completed codes are secure, and 123 contain vulnerabilities. Three of the generated code snippets are difficult to be classified as vulnerable or secure without executing the SCs and are, therefore, not categorized. The results show that 123 out of 173 (71\%) generated codes are vulnerable.

\subsection{Vulnerability labeling and model vulnerability-tuning}
\label{sec:vulnerablity-labeling}

\subsubsection{Labeled Vulnerable Smart Contracts}
\label{sec:labeled-vulnerable-smart-contracts}
The resulting distributions of the various vulnerability types in \Cref{tab:vulnerability-distribution} show that some vulnerabilities, e.g., FE, are almost non-existing. Some, e.g., RE, also have very few or non-existing testing examples.

For each of the vulnerabilities described in \cref{sec:eval-security-synthesis}, we devise the following vulnerability labels: \textless DC\textgreater, \textless IOU\textgreater, \textless NC\textgreater, \textless RE\textgreater, \textless TD\textgreater, \textless TO\textgreater, \textless TOD\textgreater, \textless UcC\textgreater, \textless UpS\textgreater, and \textless FE\textgreater. We developed scripts to insert the vulnerability labels at the relevant vulnerable lines to facilitate using the dataset of labeled SCs for \acrfull{clm} fine-tuning.

\subsubsection{Using labeled vulnerable code for vulnerability-tuning}
\label{sec:vulnerability-labeling-fine-tuning}

To perform vulnerability-tuning of the model on the labeled vulnerability dataset, we first add the vulnerability labels as special tokens to the tokenizer. This prevents the tokenizer from splitting the vulnerability labels into multiple already pre-trained tokens. For example, the ``\textless IOU\textgreater'' label is tokenized into four different tokens: ``\textless'', ``I'', ``OU'', ``\textgreater'' with corresponding ids: 27, 40, 2606, 29. 

These tokens may also be part of making up other words, which might confuse the model during training, making it challenging to identify the labels successfully. To mitigate this, the vulnerability labels are added as special tokens to the tokenizer, effectively expanding the vocabulary. We resize the model's embedding matrix to accommodate the new tokens, adding randomly initialized vectors at the end.

Starting from the fine-tuned model derived in \cref{sec:smart-contract-generation-model}, we run vulnerability-tuning of it on the dataset with the labeled vulnerabilities for two epochs. The vulnerability-tuning is completed in approximately \textbf{one hour} using \textbf{four NVIDIA A100 40GB GPUs}.

\subsubsection{Evaluate the performance of vulnerability-tuning}
\label{sec:eval-performance-vulnerability-labeling-model}

After vulnerability-tuning, we evaluate  \textbf{how
accurately the model can identify vulnerability before applying the vulnerability-constrained decoding.}

We still use the 176 vulnerability samples shown in \Cref{tab:vulnerability-distribution} as the test dataset. For each sample, we split the contract right above the vulnerable line. We then auto-complete the function using the model. The difference from the auto-completion in \cref{sec:eval-security-synthesis} is that we use the model after vulnerability-tuning to generate code in this step. After auto-completion, we manually investigate whether the generated codes are vulnerable. In addition, we count how many of the generated vulnerable codes are labeled with the correct type by the model. We only consider the immediate scope of the originally labeled vulnerability line.

The results show that codes generated from 129 of the 176 (73\%) samples are vulnerable. As the model used in this step is the one after vulnerability-tuning using the labeled samples, and there are randomnesses of the model when generating code, the number, i.e., 129, of the generated vulnerable code is slightly different from the number, i.e., 123 (see \cref{sec:eval-security-synthesis}), of the generated vulnerable code using the model without vulnerability-tuning. 

\begin{figure}[htp]
    \centering
    \input{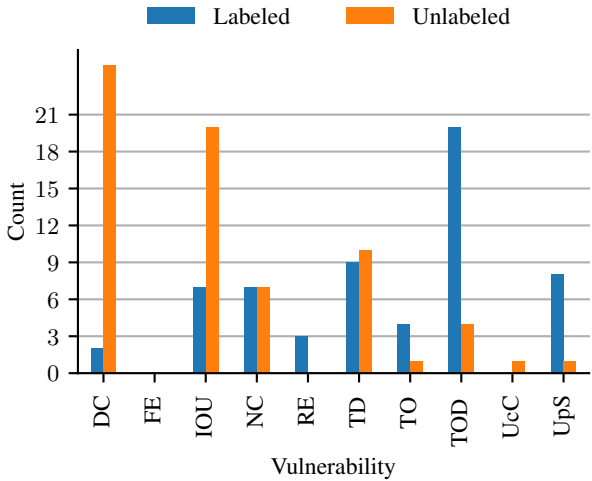}
    \caption{Correctly labeled vs incorrectly unlabeled.}
    \label{fig:labeled_vs_unlabeled}
\end{figure}

Among the 129 generated vulnerable codes, the model can label 60 of them successfully. Sixty-nine of the samples are incorrectly not labeled. The distributions of the correctly labeled and the incorrectly unlabeled samples are shown in \Cref{fig:labeled_vs_unlabeled}.

\begin{figure}[htp]
    \centering
    \input{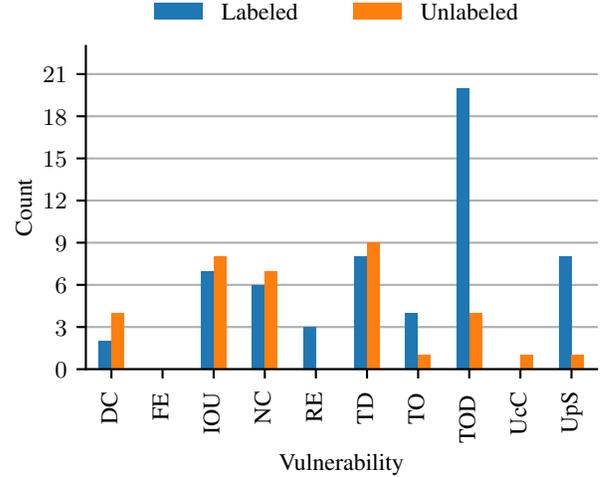}
    \caption{Correctly labeled vs incorrectly unlabeled (one vulnerability type per contract address).}
    \label{fig:unique_labeled_vs_unlabeled}
\end{figure}

After manually reviewing the samples, we discovered that some contracts contain multiple repeated occurrences of the same type of vulnerability, such as DC and IOU, as illustrated in \Cref{fig:labeled_vs_unlabeled}. We hypothesize that this is due to some training examples that have been memorized during the initial fine-tuning of the model in \cref{sec:smart-contract-generation-model}. Given our relatively small testing dataset, this would give a biased view of the model's performance. We, therefore, filter our results by uniqueness, meaning we only retain one occurrence of each vulnerability type per contract address. This greatly reduces the number of incorrectly unlabeled vulnerabilities to 35 while retaining 58 of the correctly labeled ones. This gives a success rate for correctly labeling vulnerabilities of 62\%. The distributions of the different vulnerability types for these results are shown \Cref{fig:unique_labeled_vs_unlabeled}.

\subsection{Vulnerability avoidance}
\label{sec:vulnerability-avoidance}

\subsubsection{Vulnerability-constrained decoding implementation}

\begin{figure}[htp]
    \centering
    \includegraphics[width=.75\columnwidth]{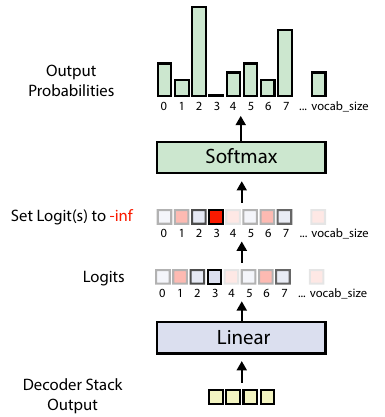}
    \caption{Vulnerability-constrained decoding implementation.}
    \label{fig:vulnerability_avoidance_implementation}
\end{figure}

To implement the vulnerability-constrained decoding, we simply set the logits of the vulnerability labels (tokens) to -infinity. As shown in the architecture overview in \Cref{fig:vulnerability_avoidance_implementation}, logits are the output of the last linear layer before going through the softmax layer. The index of the position of each logit corresponds to a specific token in the model's vocabulary. The logits can take any values from -infinity to +infinity. These are then transformed to values between zero and one by the softmax activation function. Combined, they represent the probability distribution for each token in the vocabulary. Assuming the token at index three is a vulnerability label, setting this logit to -infinity (represented in red \lstinline{-inf} in \Cref{fig:vulnerability_avoidance_implementation}) changes the probability of the next token being a vulnerability label to zero.

\subsubsection{Evaluate the performance of the vulnerability-constrained decoding}

To evaluate the performance of vulnerability-constrained decoding, the evaluation question is: \textbf{How well does the vulnerability-constrained decoding avoid generating vulnerable codes?} 

To answer this question, we follow the same approach as in \cref{sec:eval-security-synthesis}. However, we use the model after vulnerability-tuning and apply the vulnerability-constrained decoding method. We use the same samples as in \cref{sec:eval-performance-vulnerability-labeling-model}. If a sample is labeled correctly, we investigate whether the model using the vulnerability-constrained decoding approach avoided generating vulnerable codes.

\begin{figure}[htp]
    \centering
    \input{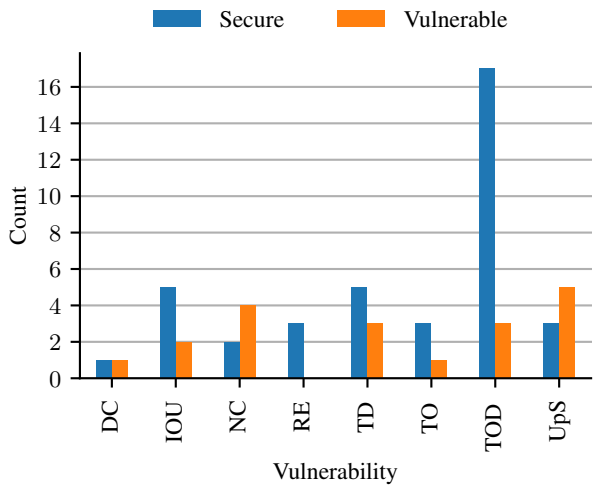}
    \caption{Secure vs vulnerable result given correctly labeled.}
    \label{fig:secure_vs_vulnerable}
\end{figure}

As described in \cref{sec:eval-performance-vulnerability-labeling-model}, the model after vulnerability-tuning could label 58 samples correctly. Of these, the vulnerability-constrained decoding method can make 39 of these secure, while 19 remain vulnerable. This means 67\% of these vulnerabilities can be avoided using our vulnerability-constrained decoding approach. The distributions of the secure versus vulnerable codes generated are shown in \Cref{fig:secure_vs_vulnerable}.

By manual evaluations of the auto-completed code, we see some emerging patterns in the secure code produced by the model. Following are a few examples in which we show the diff between the generated code without and with vulnerability-constrained decoding. A horizontal line is inserted to indicate where the generation started. One of the most common patterns is the model cannot produce ``alternative secure code''. Instead, it produces whitespace characters. Around 50\% of the examples show this characteristic. An example of a TOD vulnerability can be seen in \Cref{lst:generated-tod-example}. A TOD vulnerability often requires a rewrite of the general contract logic. In this example, the vulnerability is correctly labeled with ``\textless TOD\textgreater''. However, as the model has been provided with substantial code context (not shown), it cannot generate alternative code and produces whitespace instead. As mentioned, to generate secure code in the first place, we believe it is better not to generate code rather than to generate vulnerable code.  

\newcommand{\lstbreak}{\break \raisebox{0ex}[0ex][0ex]{\ensuremath{\color{gray}\hookrightarrow\space}}}
\begin{lstlisting}[
    caption={TOD example.},
    label=lst:generated-tod-example,
    language=diff,
    belowskip=0pt,
    escapechar=§]
   function StopGame() public payable {
       require(msg.sender==questionSender);
\end{lstlisting}
\begin{lstlisting}[
    frame=bottomline,
    firstnumber=3,
    language=diff,
    aboveskip=0pt,
    escapechar=§]
 -     <TOD>msg.sender.transfer(this.balance);
 + §\lstbg{green!40}{\space\space\space\space}§
 + §\lstbg{green!40}{\space\space\space\space}§
 + §\lstbg{green!40}{\space\space\space\space}§
   }
\end{lstlisting}

\Cref{lst:generated-TD-RE-example} shows an example where the model generated more than whitespace characters. The generated code removed the vulnerabilities but could not provide a complete function. In this example, the model identified two vulnerabilities, both a TD at line 6 and a RE at line 8. It is also successful in removing them, mainly leaving lines 10 and 12 intact. However, in the process of removing the TD vulnerability, it also removes some checks of the ``acc.balance'' state variable. If a developer wants to keep the generated line 11 as a part of the function, the developer needs to add some checks to avoid a potential IOU vulnerability.  

\begin{lstlisting}[
    caption={TD and RE example.},
    label=lst:generated-TD-RE-example,
    language=diff,
    belowskip=0pt,
    escapechar=§]
   function Collect(uint _am)
   public
   payable
   {
       var acc = Acc[msg.sender];
\end{lstlisting}
\begin{lstlisting}[
    frame=bottomline,
    firstnumber=6,
    language=diff,
    aboveskip=0pt,
    escapechar=§]
 -     <TD>if( acc.balance>=MinSum && §\lstbreak§ acc.balance>=_am && now>acc.unlockTime)
 -     {
 -         <RE>if(msg.sender.call.value(_am)())
 -         {
 -     §\lstbg{red!40}{\space\space\space\space\space\space\space\space}§acc.balance-=_am;
 +     acc.balance-=_am;
 -     §\lstbg{red!40}{\space\space\space\space\space\space\space\space}§LogFile.AddMessage(msg.sender,_am,§\lstbreak§"Collect");
 +     LogFile.AddMessage(msg.sender,_am,"Collect");
 -         }
 -     }
   }
\end{lstlisting}

\Cref{lst:generated-TD-example} shows an example that the model generated a secure solution. In this example, the TD vulnerability has been marked with the ``\textless TD\textgreater'' label in line 2 due to the \lstinline{now} keyword. Using the vulnerability-constrained decoding, the model successfully avoids this keyword and instead relies on a state variable ``refundStatus''.

\begin{lstlisting}[
    caption={TD example.},
    label=lst:generated-TD-example,
    language=diff,
    belowskip=0pt,
    escapechar=§]
   function refund () public {
\end{lstlisting}
\begin{lstlisting}[
    frame=bottomline,
    firstnumber=2,
    language=diff,
    aboveskip=0pt,
    escapechar=§]
 -     §\lstbg{red!40}{<TD>}§require(§\lstbg{red!40}{now > icoFinish}§);
 +     require(§\lstbg{green!40}{refundStatus}§);
       require(contributorBalances[msg.sender]!= 0);
       uint refundValue = contributorBalances[msg.sender];
       contributorBalances[msg.sender] = 0;
       msg.sender.transfer(refundValue);
       emit Refund(msg.sender,refundValue);
   }
\end{lstlisting}

In summary, our experiments in sections \ref{sec:eval-security-synthesis} and \ref{sec:eval-performance-vulnerability-labeling-model} show that above 70\% of the code is vulnerable. Of this 70\%, we are able to correctly label 62\%, giving us a total labeling performance of 43\%. Of this 43\%, we can avoid 67\%. This gives us a total 30\% reduction of vulnerabilities. Combined with the secure baseline percentage of 30\%, our approach can increase the amount of secure code from 30\% to 60\%.

\subsubsection{Evaluate the impact on secure code generation} After vulnerability-tuning the model using the labeled vulnerable codes, it is also important to evaluate whether our vulnerability labeling fine-tuning has negatively impacted our model for generating code in secure cases. 

To answer this question, we follow a similar evaluation procedure as that in \Cref{sec:eval-quality-synthesis}. However, the generated code now contains vulnerability labels that are not present in our original SC testing dataset. We cut out any vulnerability tokens from the generated output to calculate the BLEU score. If the model performance is unaffected, we should expect a similar BLEU score as the results of the evaluation method in \cref{sec:eval-quality-synthesis}. We get a new BLEU score of 0.544, down from 0.557. This is a small performance decrease of 2.33\%, possibly due to the random nature of the models.

\section{Related Work}
\label{chap:related-work}

\newcommand*\emptycirc[1][1ex]{\tikz\draw (0,0) circle (#1);} 
\newcommand*\halfcirc[1][1ex]{%
  \begin{tikzpicture}
  \draw[fill] (0,0)-- (90:#1) arc (90:270:#1) -- cycle ;
  \draw (0,0) circle (#1);
  \end{tikzpicture}}
\newcommand*\fullcirc[1][1ex]{\tikz\fill (0,0) circle (#1);}

One of the main problems with language models is that they often contain bias \cite{li2021detecting}, which can range from producing gender-specific jobs to favoring a certain race. In code synthesis, vulnerabilities can be considered a form of bias in language models. However, few studies focused on the security of synthesized code using transformers. Although \citet{chen2021codex} briefly discusses insecure code generated by Codex, the investigation was limited to exploring the generation of cryptography functions. While defenders can leverage new tools such as Security Copilot \cite{securitycopilot}, a Microsoft security analysis tool that enables analysts to respond to threats quickly, it is better to ensure code security in the first place, e.g., following principles like secure-by-construction \cite{securebyconstruction}.

Our approach has similarities with lexically-constrained decoding \cite{hokamp2017lexically}, which is a modification of the beam search that allows the user to specify tokens that must (or must not) appear in the decoder’s output. Post and Vilar \citet{post2018fast} proposed a variant of lexically-constrained decoding that reduced complexity from linear to constant-time, which was further optimized by \citet{hu2019improved}. While these works can deal with pre-defined constraints, our approach allows for embedding the constraint within the model through vulnerability-tuning. This enables more complex constraints, such as vulnerability patterns, to be avoided. By incorporating vulnerability-tuning, our approach can adapt and respond to changing constraints in real-time, offering enhanced flexibility and effectiveness. Furthermore, our implementation is based on a simple greedy search. We expect even better results using more complex decoding algorithms, such as beam-search, which are left for future investigations.

\Acrfull{rl} has been successfully used to align models according to human preferences \cite{ouyang2022training}. It is also known as \acrfull{rlhf} and powers ChatGPT \cite{chatgpt}. RL has also been applied to program synthesis. For example, \cite{le2022coderl} studied the integration of \acrshort{rl} with unit test signals in the fine-tuning of program synthesis models, and \cite{shojaee2023executionbased} further expanded on this idea, creating an \acrshort{rl} framework applicable to various code generation tasks and programming languages by employing execution feedback as an external source of knowledge in model optimization. Although \acrshort{rl} could be a strong candidate for reducing code vulnerabilities, it is a rather slow and resource-intensive process \cite{janner2021offline}. Unlike the \acrshort{rl}-based model fine-tuning, our approach prioritizes fast model updates to instantly mitigate the risk of generating vulnerable code. People can use our approach for quick and minor model updates. The slow and resource-intensive approaches, e.g., retraining the entire model using only secure code or the RLHF approach, are probably more suitable for major model updates.

\section{Discussion}
\label{chap:discussion}

In this section, we discuss the implications of our results and the threats to the validity of our study.

\subsection{Implication to academia and industry}
\label{sec:rq1-implication-to-academia-and-industry}

Vulnerabilities can be considered as a form of bias in language models. Hence, the vulnerability-constrained decoding approach could be generalized to handle other types of biases, such as gender bias. Since the approach relies primarily on augmenting the dataset, the method should also be transferable to other models. 

Existing code synthesis solutions based on transformers may produce many vulnerabilities \citet{pearch2021asleep}. Reducing vulnerabilities could bring significant benefits to the industry. Our results show that it is possible to quickly and incrementally fine-tune the code generator using information related to the vulnerabilities in the training dataset that are often gradually identified after the model is trained. This opens the possibilities for continuous model improvements to generate secure code. In this study, using only 941 labeled vulnerabilities to fine-tune the model within one hour has already significantly improved the model's capability to generate secure code. We believe fine-tuning the model with more vulnerability samples could further improve its performance.

The transformer model fine-tuned for SC code generation can accurately generate SC code with high BLEU and CrystalBLEU scores. Using this model in an industry setting can greatly reduce the efforts needed for creating SCs. Although Decrypted \cite{Decrypted} uses large-scale language models to generate SCs, it is in beta (since June 2021) and is close-sourced. There are no details about its model, training data, etc. 

A side product of the fine-tuned model is the construction of the currently largest dataset of real SCs, consisting of 186,397 SCs. The largest competing dataset \cite{ren2021empirical} contains 45,622 real-world SCs, filtered from 1.5 million contracts by comparing the MD5 hash of the contracts. However, their data are unfit for use in deep learning applications because they did not inflate the contracts to remove library code.

\subsection{Threats to validity}
\label{sec:rq1-threats-to-validity}

Many of the SCs contain duplicated code. If these duplications make up too large of a percentage of the dataset, this will lead to overly optimistic estimates of the model's performance. We performed an extensive filter of code duplications. Thus, we believe that the risk of cross-contamination in our study is relatively low.

A potential threat is that the manual classifications of the secure and vulnerable code may be biased. To mitigate the threat, all classifications of secure and vulnerable code are performed and cross-validated by two researchers. Another potential threat is the quality of the labeled data. Although we used the dataset of \cite{Hu2023}, which includes vulnerabilities that are verified by multiple code analysis tools with manual investigations, there are always false negatives and positives in the dataset, which may slightly impact our results. To evaluate the quality of the generated code, we use the BLEU and CrystalBLEU scores and leave alternative evaluation methods for SC code synthesis for future research.  

Concerning external validity, the approach does not rely on the specifics of SCs, other programming languages, or LLM models. However, since SCs are smaller and less complex than applications developed using other programming languages, there may be some restrictions on the generalizability of the approach.

\section{Conclusion and Future Work}
\label{chap:conclusion}

To address the issue that state-of-the-art LLMs generate vulnerabilities during code completion, we proposed a novel vulnerability-constrained decoding approach to drive the model to generate secure code. Our results provide the first empirical evidence that using static analysis tools to label vulnerable code and use them to fine-tune the model with the application of the vulnerability-constrained decoding approach can efficiently and effectively reduce the possibility of generating vulnerable code.

In the future, we plan to fine-tune the model with more vulnerable code samples to improve its performance. In addition, we want to investigate the possibility of using the model to label and auto-repair the vulnerable code typed in by the developers.

\section{Data Availability}
\label{chap:data-availability}

We make all the SC datasets publicly available.
\begin{itemize}
    \item Verified Smart Contracts dataset:\\ \url{https://doi.org/10.6084/m9.figshare.20781799}
    \item Verified Smart Contract Code Comments dataset:\\ \url{https://doi.org/10.6084/m9.figshare.20780878}
    \item Vulnerable Verified Smart Contracts dataset:\\ \url{https://doi.org/10.6084/m9.figshare.21990287}
\end{itemize}

The final version of the SC model is also released.
\begin{itemize}
    \item Vulnerability-tuned GPT-J 6B model:\\ \url{https://doi.org/10.5281/zenodo.7595802}
\end{itemize}

\section*{Acknowledgments}
This work is jointly supported by the National Key Research and Development Program of China (No. 2019YFE0105500) and the Research Council of Norway (No. 309494) and the Key Research and Development Program of Jiangsu
Province (No. BE2021002-3). Tianyuan Hu thanks the Chinese Scholarship Council (CSC) for financial support (202106090057).

\bibliographystyle{IEEEtran}
\bibliography{references}

\begin{thebibliography}{10}
\providecommand{\url}[1]{#1}
\csname url@samestyle\endcsname
\providecommand{\newblock}{\relax}
\providecommand{\bibinfo}[2]{#2}
\providecommand{\BIBentrySTDinterwordspacing}{\spaceskip=0pt\relax}
\providecommand{\BIBentryALTinterwordstretchfactor}{4}
\providecommand{\BIBentryALTinterwordspacing}{\spaceskip=\fontdimen2\font plus
\BIBentryALTinterwordstretchfactor\fontdimen3\font minus
  \fontdimen4\font\relax}
\providecommand{\BIBforeignlanguage}[2]{{%
\expandafter\ifx\csname l@#1\endcsname\relax
\typeout{** WARNING: IEEEtran.bst: No hyphenation pattern has been}%
\typeout{** loaded for the language `#1'. Using the pattern for}%
\typeout{** the default language instead.}%
\else
\language=\csname l@#1\endcsname
\fi
#2}}
\providecommand{\BIBdecl}{\relax}
\BIBdecl

\bibitem{alon2018code2vec}
\BIBentryALTinterwordspacing
U.~Alon, M.~Zilberstein, O.~Levy, and E.~Yahav, ``code2vec: Learning
  distributed representations of code,'' 2018. [Online]. Available:
  \url{https://arxiv.org/abs/1803.09473}
\BIBentrySTDinterwordspacing

\bibitem{iyer2018mapping}
\BIBentryALTinterwordspacing
S.~Iyer, I.~Konstas, A.~Cheung, and L.~Zettlemoyer, ``Mapping language to code
  in programmatic context,'' 2018. [Online]. Available:
  \url{https://arxiv.org/abs/1808.09588}
\BIBentrySTDinterwordspacing

\bibitem{peters2018deep}
\BIBentryALTinterwordspacing
M.~E. Peters, M.~Neumann, M.~Iyyer, M.~Gardner, C.~Clark, K.~Lee, and
  L.~Zettlemoyer, ``Deep contextualized word representations,'' 2018. [Online].
  Available: \url{https://arxiv.org/abs/1802.05365}
\BIBentrySTDinterwordspacing

\bibitem{radford2018improving}
A.~Radford, K.~Narasimhan, T.~Salimans, and I.~Sutskever, ``Improving language
  understanding by generative pre-training,'' 2018.

\bibitem{copilot}
\BIBentryALTinterwordspacing
GitHub. (2022) Your ai pair programmer. [Online]. Available:
  \url{https://github.com/features/copilot}
\BIBentrySTDinterwordspacing

\bibitem{alphacode}
\BIBentryALTinterwordspacing
Y.~Li, D.~Choi, J.~Chung, N.~Kushman, J.~Schrittwieser, R.~Leblond, T.~Eccles,
  J.~Keeling, F.~Gimeno, A.~D. Lago, T.~Hubert, P.~Choy, C.~d.~M. d'Autume,
  I.~Babuschkin, X.~Chen, P.-S. Huang, J.~Welbl, S.~Gowal, A.~Cherepanov,
  J.~Molloy, D.~J. Mankowitz, E.~S. Robson, P.~Kohli, N.~de~Freitas,
  K.~Kavukcuoglu, and O.~Vinyals, ``Competition-level code generation with
  alphacode,'' 2022. [Online]. Available:
  \url{https://arxiv.org/abs/2203.07814}
\BIBentrySTDinterwordspacing

\bibitem{christopoulou2022pangu}
F.~Christopoulou, G.~Lampouras, M.~Gritta, G.~Zhang, Y.~Guo, Z.~Li, Q.~Zhang,
  M.~Xiao, B.~Shen, L.~Li \emph{et~al.}, ``Pangu-coder: Program synthesis with
  function-level language modeling,'' \emph{arXiv preprint arXiv:2207.11280},
  2022.

\bibitem{chen2021codex}
\BIBentryALTinterwordspacing
M.~Chen, J.~Tworek, H.~Jun, Q.~Yuan, H.~P. d.~O. Pinto, J.~Kaplan, H.~Edwards,
  Y.~Burda, N.~Joseph, G.~Brockman, A.~Ray, R.~Puri, G.~Krueger, M.~Petrov,
  H.~Khlaaf, G.~Sastry, P.~Mishkin, B.~Chan, S.~Gray, N.~Ryder, M.~Pavlov,
  A.~Power, L.~Kaiser, M.~Bavarian, C.~Winter, P.~Tillet, F.~P. Such,
  D.~Cummings, M.~Plappert, F.~Chantzis, E.~Barnes, A.~Herbert-Voss, W.~H.
  Guss, A.~Nichol, A.~Paino, N.~Tezak, J.~Tang, I.~Babuschkin, S.~Balaji,
  S.~Jain, W.~Saunders, C.~Hesse, A.~N. Carr, J.~Leike, J.~Achiam, V.~Misra,
  E.~Morikawa, A.~Radford, M.~Knight, M.~Brundage, M.~Murati, K.~Mayer,
  P.~Welinder, B.~McGrew, D.~Amodei, S.~McCandlish, I.~Sutskever, and
  W.~Zaremba, ``Evaluating large language models trained on code,'' 2021.
  [Online]. Available: \url{https://arxiv.org/abs/2107.03374}
\BIBentrySTDinterwordspacing

\bibitem{chatgpt}
\BIBentryALTinterwordspacing
OpenAI. (2022, 1) Introducing chatgpt. [Online]. Available:
  \url{https://openai.com/blog/chatgpt}
\BIBentrySTDinterwordspacing

\bibitem{derner2023safeguards}
E.~Derner and K.~Batistič, ``Beyond the safeguards: Exploring the security
  risks of chatgpt,'' 2023.

\bibitem{pearce2021asleep}
\BIBentryALTinterwordspacing
H.~Pearce, B.~Ahmad, B.~Tan, B.~Dolan-Gavitt, and R.~Karri, ``Asleep at the
  keyboard? assessing the security of github copilot's code contributions,''
  2021. [Online]. Available: \url{https://arxiv.org/abs/2108.09293}
\BIBentrySTDinterwordspacing

\bibitem{khoury2023secure}
R.~Khoury, A.~R. Avila, J.~Brunelle, and B.~M. Camara, ``How secure is code
  generated by chatgpt?'' 2023.

\bibitem{atzei2017survey}
N.~Atzei, M.~Bartoletti, and T.~Cimoli, ``A survey of attacks on ethereum smart
  contracts (sok),'' in \emph{International conference on principles of
  security and trust}.\hskip 1em plus 0.5em minus 0.4em\relax Springer, 2017,
  pp. 164--186.

\bibitem{gpt-j}
B.~Wang and A.~Komatsuzaki, ``{GPT-J-6B: A 6 Billion Parameter Autoregressive
  Language Model},'' \url{https://github.com/kingoflolz/mesh-transformer-jax},
  May 2021.

\bibitem{Bleuscore}
\BIBentryALTinterwordspacing
(2021) Foundations of nlp explained. [Online]. Available:
  \url{https://towardsdatascience.com/foundations-of-nlp-explained-bleu-score-and-wer-metrics-1a5ba06d812b.}
\BIBentrySTDinterwordspacing

\bibitem{Hu2023}
\BIBentryALTinterwordspacing
T.~Hu, J.~Li, B.~Li, and A.~Storhaug, ``{Why Smart Contracts Reported as
  Vulnerable were not Exploited?}'' 3 2023. [Online]. Available:
  \url{https://www.techrxiv.org/articles/preprint/Why_Smart_Contracts_Reported_as_Vulnerable_were_not_Exploited_/21953189}
\BIBentrySTDinterwordspacing

\bibitem{oyente2016making}
L.~Luu, D.-H. Chu, H.~Olickel, P.~Saxena, and A.~Hobor,
  ``\BIBforeignlanguage{{English}}{{Making Smart Contracts Smarter}},'' in
  \emph{\BIBforeignlanguage{{English}}{{CCS'16: PROCEEDINGS OF THE 2016 ACM
  SIGSAC CONFERENCE ON COMPUTER AND COMMUNICATIONS SECURITY}}}, {Assoc Comp
  Machinery; ACM Special Interest Grp Secur Audit \& Control}.\hskip 1em plus
  0.5em minus 0.4em\relax {1515 BROADWAY, NEW YORK, NY 10036-9998 USA}: {ASSOC
  COMPUTING MACHINERY}, 2016, {Proceedings Paper}, pp. {254--269}, {23rd ACM
  Conference on Computer and Communications Security (CCS), Vienna, AUSTRIA,
  OCT 24-28, 2016}.

\bibitem{feist2019slither}
J.~Feist, G.~Grieco, and A.~Groce, ``Slither: a static analysis framework for
  smart contracts,'' in \emph{2019 IEEE/ACM 2nd International Workshop on
  Emerging Trends in Software Engineering for Blockchain (WETSEB)}.\hskip 1em
  plus 0.5em minus 0.4em\relax IEEE, 2019, pp. 8--15.

\bibitem{mythril}
\BIBentryALTinterwordspacing
(2018) Mythril: an open-source security analysis tool for ethereum smart
  contracts. [Online]. Available: \url{https://github.com/ConsenSys/mythril}
\BIBentrySTDinterwordspacing

\bibitem{ContractFuzzer}
\BIBentryALTinterwordspacing
B.~Jiang, Y.~Liu, and W.~K. Chan, ``Contractfuzzer: Fuzzing smart contracts for
  vulnerability detection,'' ser. ASE '18.\hskip 1em plus 0.5em minus
  0.4em\relax New York, NY, USA: Association for Computing Machinery, 2018, p.
  259–269. [Online]. Available: \url{https://doi.org/10.1145/3238147.3238177}
\BIBentrySTDinterwordspacing

\bibitem{sFuzz}
\BIBentryALTinterwordspacing
T.~D. Nguyen, L.~H. Pham, J.~Sun, Y.~Lin, and Q.~T. Minh, ``Sfuzz: An efficient
  adaptive fuzzer for solidity smart contracts,'' in \emph{Proceedings of the
  ACM/IEEE 42nd International Conference on Software Engineering}, ser. ICSE
  '20.\hskip 1em plus 0.5em minus 0.4em\relax New York, NY, USA: Association
  for Computing Machinery, 2020, p. 778–788. [Online]. Available:
  \url{https://doi.org/10.1145/3377811.3380334}
\BIBentrySTDinterwordspacing

\bibitem{SMARTIAN}
J.~Choi, D.~Kim, S.~Kim, G.~Grieco, A.~Groce, and S.~K. Cha, ``Smartian:
  Enhancing smart contract fuzzing with static and dynamic data-flow
  analyses,'' in \emph{2021 36th IEEE/ACM International Conference on Automated
  Software Engineering (ASE)}, 2021, pp. 227--239.

\bibitem{allamanis_adverse_2019}
M.~Allamanis, ``The adverse effects of code duplication in machine learning
  models of code,'' in \emph{Proceedings of the 2019 {ACM} {SIGPLAN}
  {International} {Symposium} on {New} {Ideas}, {New} {Paradigms}, and
  {Reflections} on {Programming} and {Software}}, 2019, pp. 143--153.

\bibitem{jaccard}
\BIBentryALTinterwordspacing
DeepAi. (2022, 4) What is the jaccard index? [Online]. Available:
  \url{https://deepai.org/machine-learning-glossary-and-terms/jaccard-index}
\BIBentrySTDinterwordspacing

\bibitem{gao2021thepile}
\BIBentryALTinterwordspacing
L.~Gao, S.~Biderman, S.~Black, L.~Golding, T.~Hoppe, C.~Foster, J.~Phang,
  H.~He, A.~Thite, N.~Nabeshima, S.~Presser, and C.~Leahy, ``The pile: An 800gb
  dataset of diverse text for language modeling,'' 2021. [Online]. Available:
  \url{https://arxiv.org/abs/2101.00027}
\BIBentrySTDinterwordspacing

\bibitem{eleutherai}
\BIBentryALTinterwordspacing
EleutherAI. (2022, 4) Eleutherai. [Online]. Available:
  \url{https://www.eleuther.ai}
\BIBentrySTDinterwordspacing

\bibitem{wolf2021}
\BIBentryALTinterwordspacing
M.~Woolf. (2021, 6) Fun and dystopia with ai-based code generation using
  gpt-j-6b. [Online]. Available: \url{https://minimaxir.com/2021/06/gpt-j-6b/}
\BIBentrySTDinterwordspacing

\bibitem{su2021roformer}
\BIBentryALTinterwordspacing
J.~Su, Y.~Lu, S.~Pan, B.~Wen, and Y.~Liu, ``Roformer: Enhanced transformer with
  rotary position embedding,'' 2021. [Online]. Available:
  \url{https://arxiv.org/abs/2104.09864}
\BIBentrySTDinterwordspacing

\bibitem{solidifi}
A.~Ghaleb and K.~Pattabiraman, ``How effective are smart contract analysis
  tools? evaluating smart contract static analysis tools using bug injection,''
  in \emph{ISSTA '20: 29th ACM SIGSOFT International Symposium on Software
  Testing and Analysis}, 2020.

\bibitem{EmpiricalOnSmartBug}
\BIBentryALTinterwordspacing
T.~Durieux, J.~a.~F. Ferreira, R.~Abreu, and P.~Cruz, ``Empirical review of
  automated analysis tools on 47,587 ethereum smart contracts,'' in
  \emph{Proceedings of the ACM/IEEE 42nd International Conference on Software
  Engineering}, ser. ICSE'20.\hskip 1em plus 0.5em minus 0.4em\relax New York,
  NY, USA: Association for Computing Machinery, 2020, pp. 530--541. [Online].
  Available: \url{https://doi.org/10.1145/3377811.3380364}
\BIBentrySTDinterwordspacing

\bibitem{Freitag_2017}
\BIBentryALTinterwordspacing
M.~Freitag and Y.~Al-Onaizan, ``Beam search strategies for neural machine
  translation,'' in \emph{Proceedings of the First Workshop on Neural Machine
  Translation}.\hskip 1em plus 0.5em minus 0.4em\relax Association for
  Computational Linguistics, 2017. [Online]. Available:
  \url{https://doi.org/10.18653/v1/w17-3207}
\BIBentrySTDinterwordspacing

\bibitem{deepspeed}
\BIBentryALTinterwordspacing
Microsoft, ``Deepspeed,'' 2022. [Online]. Available:
  \url{https://www.deepspeed.ai/}
\BIBentrySTDinterwordspacing

\bibitem{ren2020codebleu}
\BIBentryALTinterwordspacing
S.~Ren, D.~Guo, S.~Lu, L.~Zhou, S.~Liu, D.~Tang, N.~Sundaresan, M.~Zhou,
  A.~Blanco, and S.~Ma, ``Codebleu: a method for automatic evaluation of code
  synthesis,'' 2020. [Online]. Available:
  \url{https://arxiv.org/abs/2009.10297}
\BIBentrySTDinterwordspacing

\bibitem{CrystalBLEU}
\BIBentryALTinterwordspacing
A.~Eghbali and M.~Pradel, ``Crystalbleu: Precisely and efficiently measuring
  the similarity of code,'' in \emph{Proceedings of the 37th IEEE/ACM
  International Conference on Automated Software Engineering}, ser. ASE
  '22.\hskip 1em plus 0.5em minus 0.4em\relax New York, NY, USA: Association
  for Computing Machinery, 2023. [Online]. Available:
  \url{https://doi.org/10.1145/3551349.3556903}
\BIBentrySTDinterwordspacing

\bibitem{safemath}
\BIBentryALTinterwordspacing
SafeMath, ``Safemath,'' 2022. [Online]. Available:
  \url{https://docs.openzeppelin.com/contracts/3.x/utilities#math}
\BIBentrySTDinterwordspacing

\bibitem{li2021detecting}
\BIBentryALTinterwordspacing
B.~Li, H.~Peng, R.~Sainju, J.~Yang, L.~Yang, Y.~Liang, W.~Jiang, B.~Wang,
  H.~Liu, and C.~Ding, ``Detecting gender bias in transformer-based models: A
  case study on bert,'' 2021. [Online]. Available:
  \url{https://arxiv.org/abs/2110.15733}
\BIBentrySTDinterwordspacing

\bibitem{securitycopilot}
\BIBentryALTinterwordspacing
Microsoft. (2023, 3) Introducing microsoft security copilot: Empowering
  defenders at the speed of ai. [Online]. Available:
  \url{https://blogs.microsoft.com/blog/2023/03/28/introducing-microsoft-security-copilot-empowering-defenders-at-the-speed-of-ai/}
\BIBentrySTDinterwordspacing

\bibitem{securebyconstruction}
S.~Liu, A.~Trivedi, X.~Yin, and M.~Zamani, ``Secure-by-construction synthesis
  of cyber-physical systems,'' 2022.

\bibitem{hokamp2017lexically}
\BIBentryALTinterwordspacing
C.~Hokamp and Q.~Liu, ``Lexically constrained decoding for sequence generation
  using grid beam search,'' in \emph{Proceedings of the 55th Annual Meeting of
  the Association for Computational Linguistics (Volume 1: Long Papers)}.\hskip
  1em plus 0.5em minus 0.4em\relax Vancouver, Canada: Association for
  Computational Linguistics, Jul. 2017, pp. 1535--1546. [Online]. Available:
  \url{https://aclanthology.org/P17-1141}
\BIBentrySTDinterwordspacing

\bibitem{post2018fast}
\BIBentryALTinterwordspacing
M.~Post and D.~Vilar, ``Fast lexically constrained decoding with dynamic beam
  allocation for neural machine translation,'' in \emph{Proceedings of the 2018
  Conference of the North {A}merican Chapter of the Association for
  Computational Linguistics: Human Language Technologies, Volume 1 (Long
  Papers)}.\hskip 1em plus 0.5em minus 0.4em\relax New Orleans, Louisiana:
  Association for Computational Linguistics, Jun. 2018, pp. 1314--1324.
  [Online]. Available: \url{https://aclanthology.org/N18-1119}
\BIBentrySTDinterwordspacing

\bibitem{hu2019improved}
\BIBentryALTinterwordspacing
J.~E. Hu, H.~Khayrallah, R.~Culkin, P.~Xia, T.~Chen, M.~Post, and B.~Van~Durme,
  ``Improved lexically constrained decoding for translation and monolingual
  rewriting,'' in \emph{Proceedings of the 2019 Conference of the North
  {A}merican Chapter of the Association for Computational Linguistics: Human
  Language Technologies, Volume 1 (Long and Short Papers)}.\hskip 1em plus
  0.5em minus 0.4em\relax Minneapolis, Minnesota: Association for Computational
  Linguistics, Jun. 2019, pp. 839--850. [Online]. Available:
  \url{https://aclanthology.org/N19-1090}
\BIBentrySTDinterwordspacing

\bibitem{ouyang2022training}
L.~Ouyang, J.~Wu, X.~Jiang, D.~Almeida, C.~L. Wainwright, P.~Mishkin, C.~Zhang,
  S.~Agarwal, K.~Slama, A.~Ray, J.~Schulman, J.~Hilton, F.~Kelton, L.~Miller,
  M.~Simens, A.~Askell, P.~Welinder, P.~Christiano, J.~Leike, and R.~Lowe,
  ``Training language models to follow instructions with human feedback,''
  2022.

\bibitem{le2022coderl}
H.~Le, Y.~Wang, A.~D. Gotmare, S.~Savarese, and S.~C.~H. Hoi, ``Coderl:
  Mastering code generation through pretrained models and deep reinforcement
  learning,'' 2022.

\bibitem{shojaee2023executionbased}
P.~Shojaee, A.~Jain, S.~Tipirneni, and C.~K. Reddy, ``Execution-based code
  generation using deep reinforcement learning,'' 2023.

\bibitem{janner2021offline}
M.~Janner, Q.~Li, and S.~Levine, ``Offline reinforcement learning as one big
  sequence modeling problem,'' 2021.

\bibitem{pearch2021asleep}
\BIBentryALTinterwordspacing
H.~Pearce, B.~Ahmad, B.~Tan, B.~Dolan-Gavitt, and R.~Karri, ``Asleep at the
  keyboard? assessing the security of github copilot's code contributions,''
  2021. [Online]. Available: \url{https://arxiv.org/abs/2108.09293}
\BIBentrySTDinterwordspacing

\bibitem{Decrypted}
\BIBentryALTinterwordspacing
(2022) Decrypted. [Online]. Available:
  \url{https://decrypted-ai.chronologic.network/)}
\BIBentrySTDinterwordspacing

\bibitem{ren2021empirical}
\BIBentryALTinterwordspacing
M.~Ren, Z.~Yin, F.~Ma, Z.~Xu, Y.~Jiang, C.~Sun, H.~Li, and Y.~Cai, ``Empirical
  evaluation of smart contract testing: What is the best choice?'' in
  \emph{Proceedings of the 30th ACM SIGSOFT International Symposium on Software
  Testing and Analysis}, ser. ISSTA 2021.\hskip 1em plus 0.5em minus
  0.4em\relax New York, NY, USA: Association for Computing Machinery, 2021, p.
  566–579. [Online]. Available: \url{https://doi.org/10.1145/3460319.3464837}
\BIBentrySTDinterwordspacing

\end{thebibliography}

\end{document}